\documentstyle[prd,tighten,aps]{revtex}
\begin{document}
\draft
\twocolumn[\hsize\textwidth\columnwidth\hsize\csname
@twocolumnfalse\endcsname

\title{ Symmetries and degrees of freedom in 2-dimensional dual models }
\author{C. P. Constantinidis$^*$ and F. P. Devecchi$^{**}$}
\address{$^*$ Depto. de F\'{\i}sica, Universidade Federal do Esp\'{\i}rito
 Santo,
av. F. Ferrari-Goiabeiras,\\  cep 29060.900, Vit\'oria-ES, Brazil.}
\address{$^{**}$ Depto. de F\'{\i}sica, Universidade Federal do Paran\'a,
c.p. 19091, cep 81531.990, Curitiba-PR, Brazil.}
\maketitle 
\begin{abstract} 
The 2-dimensional version of the Schwarz and Sen duality model (Tseytlin model)
is analised at the classical and quantum levels. The solutions are obtained 
 after removing
the gauge dependent sector using the Dirac method. The Poincar\`e invariance is
 verified at both levels. An extension with global supersymmetry is also 
proposed.
\end{abstract}
  

\vskip2pc]  

The 4-dimensional Schwarz-Sen duality model\cite{Sen} has received much
attention in the
 last few
years\cite{Peo} . In \cite{Sen} one remark was that a 2-dimensional
 version of
 this model
was exactly Tseytlin model\cite{Tsey}, and that
 both 2
 and 4-dimensional versions
were related to the formulation of free P-forms \cite{Hen} \cite{Hen1}
 .

In a two dimensional space-time
duality  is a discrete symmetry\cite{Hen1}. As a consequence Tseytlin model
has a classically canonical relation to Floreanini-Jackiw self-dual bosons
 model
 (FJ)\cite{Flo}\cite{Sen}.
 This formulation was  studied in\cite{Dev} for  a
  compact space-time, obtaining carefully its constraints structure
  together with the classical and quantum solutions. 
The Tseytlin model can be studied, at both classical and quantum levels, 
 with the same techiques used in\cite{Dev} and   
to make this analysis is the main purpose of this work. A supersymmetric
extension is also constructed.
\par \par 
We start our analysis  with the
 Tseytlin action\cite{Tsey}, whose dynamics is described by the
non-covariant Lagrangian density

\begin{equation}
{\cal L} = \frac{1}{2} \left[ \partial_{0} \phi ^1 \partial_{1} \phi ^2+
\partial_{0} \phi ^2 \partial_{1} \phi ^1 - \partial_{1} \phi ^1 
\partial_{1} \phi ^1 - \partial_{1} \phi ^2 
\partial_{1} \phi ^2 \right] , \label{1}
\end{equation}
  $\phi ^{\alpha }\equiv \phi ^{\alpha} (x^0,x^1)$ ,\,\,  $\alpha =1,2$, are
 real
 scalar
fields in a 2-dimensional space-time. This model was analised by Tseytlin
\cite{Tsey} in the closed string context, using functional quantization
 methods.

The canonical Hamiltonian correspondent to (\ref{1}) can be easily constructed
\begin{equation}
\int_{-\infty}^{+\infty} dx^{1} {\cal H}_c=H_{c} = \frac{1}{2} 
\int_{-\infty}^{+\infty} dx^{1} \left[\partial_{1}
\phi ^1
\partial_{1}\phi ^1+\partial_{1}\phi ^2
\partial_{1}\phi ^2\right], \label{2}
\end{equation}
and using  the Dirac formalism\cite{Dir} we can obtain the primary constraints
 structure
 
\begin{equation}
\omega ^{\alpha } (x^{0},x^{1}) = \pi ^{\alpha}(x^{0},x^{1}) - 
\frac{1}{2} 
M ^{\alpha \beta}\partial_{1}\phi ^{\beta }(x^{0},x^{1}) \approx 0\,\,\, ,
 \label{3}
\end{equation}
where $\pi ^{\alpha}$ are the momenta conjugate to $\phi ^{\alpha}$ and
$M ^{\alpha \beta}$ is  the ``duality'' matrix

\begin{eqnarray}
M ^{\alpha \beta}=\left(
\begin{array}{cc}
0 & 1 \\
1 & 0
\end{array}
\right)\label{4}
\end{eqnarray}

Using (\ref{3}) we can construct   the constraints
matrix $( \Delta )$

\begin{eqnarray}
\Delta (x^{0};x^{1},y^{1}) = \left(
 \begin{array}{ccc}
 0&-\delta ' (x-y)\\
-\delta ' (x-y)&0
\end{array}
\right),\label{5}
\end{eqnarray}        

whose inverse is not unique, this meaning 
that there is a first class sector in the constraints set, so the
 solutions  for the Euler-Lagrange equations
 \begin{equation}
\partial _1
(\partial _0\phi _1 -\partial _1 \phi _2) =0\,\,\,\,\,\,
\partial _1(\partial _0\phi _2 -
\partial _1 \phi _1)=0 \,\,\, ,\label{6}
\end{equation}
may be gauge dependent.

To clarify this point we realise that 
analogously to the case of the FJ model  the set of  constraints
 $\omega ^a (x)$ is improper\cite{Bg}, that is, the
  variation $\delta \omega ^{\alpha}$

\begin{equation}
\label{7}
\delta \omega ^{\alpha} [\eta] = \int_{- \infty}^{+ \infty} dx^{1} \left( A
^{\alpha}
(x^{1}) 
\,\delta \phi ^{\alpha} (x^{1})\,\,+\,\,B ^{\alpha}(x^{1})\,\delta 
\pi ^{\alpha}(x^{1})
 \right),
\end{equation}
with  $A (x^{1}) \equiv \delta \gamma [\eta] / \delta \phi (x^{1})$,
 $B (x^{1}) \equiv \delta \gamma [\eta] / \delta \pi (x^{1})$
and $\eta $ being the dual space functions\cite {Bg},
involves a surface term. As a consequence, the 
  choice of the boundary  conditions is a key ingredient in the classical
  analysis \cite{Dev} in the sense that  to each set of boundary
   conditions
  corresponds a different constraint
structure. 

With this in mind we analise the model with periodic
 boundary conditions, settled in a compact domain.
In this case it is possible to construct a discrete formulation,
 and the Fourier decomposition of the fields $\phi ^{\alpha}$ is 
\begin{eqnarray}
\label{8}
\phi ^{\alpha}(x^{0},x^{1})\,=\,\frac{1}{2R} a_{0}^{\alpha}(x^{0})\,\, + 
\,\,\frac{1}{2R}
\sum_{n>0}\left\{ \left[a_{n}^{\alpha}(x^{0})\,\, \right. \right. \nonumber \\
\left. + \,\,i\,
b_{n}^{\alpha}(x^{0}) \right]\,
e^{\frac{in\pi } {R} x^{1}}  \nonumber \\
\left.  +  \left[a_{n}^{\alpha}(x^{0})\,\, -
\,\,i\,b_{n}^{\alpha}(x^{0}) \right]\, e^{- \frac{in\pi } {R} x^{1}} \right\},
\end{eqnarray}
 by using (\ref{1}) and (\ref{6}) one finds for
the Lagrangian 
\begin{eqnarray}
\label{9}
L\,\equiv \, \int_{-R}^{+R} dx^{1} {\cal L}\,\nonumber \\
=\,\frac{1}{2R} \sum_{n>0}
\left[M ^{\alpha \beta}
 \left( a_{n}^{\alpha} \dot{b_{n}}^{\beta}\,-\,\dot{a_{n}}^{\alpha}
  b_{n}^{\beta}\right)\omega _n\,\,   \right. \nonumber  \\ -\,\,
 \omega_{n}^{2}
\left. \left((a_{n}^{\alpha})^{2}\,+\,(b_{n}^{\alpha})^{2}\right) \right],
\end{eqnarray}
 where the dots represent the derivatives in
 $x^{0} $ and $ \omega_{n} \equiv n\pi /R $. Following the Dirac procedure
 \cite{Dir}
  we obtain a set of 
  two  first--class constraints
\begin{equation}
\label{10}
p_{a_0}^{\alpha } \approx 0 \,\,\,\,\,\,\,\,\,\,\,\, \alpha =1,2\,\,\,\, ,
\end{equation}
and a set of primary second-class constraints ($ n > 0 $)
\begin{equation}
\label{11}
\Gamma_{p_{n}}^{\alpha +}\,\equiv\,p_{a_{n}}^{\alpha}\,+ \,
\frac{\omega_{n}}{2R}M ^{\alpha \beta} b_{n}^{\beta}
\approx 0,\,\,\,\,\,\,\, 
\end{equation}

\begin{equation}
\label{12}
\Gamma_{p_{n}}^{\alpha -}\,\equiv\,p_{b_{n}}^{\alpha}\,- \,
\frac{\omega_{n}}{2R} M ^{\alpha \beta} a_{n}^{\beta}
\approx 0,\,\,\,\,\,\,\, 
\end{equation}
where $ p_{a_{n}}^{\alpha}$ and $ p_{b_{n}}^{\alpha} $ are the canonical
 momenta
 conjugate to
$ a_{n}$ and $ b_{n}$, respectively. Furthermore, the canonical
Hamiltonian $ H_{c}^{P} $ reads
\begin{equation}
\label{13}
H_{c}^{P}\,=\,\frac{1}{2R} \sum_{n > 0} \omega_{n}^{2}
\left[({a_{n}^{\alpha}})^{2}\,+\,({b_{n}^{\alpha})^{2}}\right],
\end{equation}
and it is easy to verify that there are no secondary constraints.

 The first-class
constraints (\ref{10}) generate the following gauge
transformations
\begin{equation} 
\phi ^{\alpha}(x^{0},x^{1}) \longrightarrow \phi ^{\alpha}(x^{0},x^{1})
\,+\,f^{\alpha}(x^{0}), \label{14}
\end{equation}
that leave the action (\ref{1}) invariant. This gauge freedom can be fixed
 using a set of  external conditions that turn
the first class constraints  (\ref{10}) into second class. For our purposes
it is enough to choose

\begin{equation}
\label{15}
a_{0}^{\alpha } +  \rho ^{\alpha } \approx 0\,\,\,.
\end{equation}
where $\rho ^{\alpha } $ is an arbitrary function of the phase-space variables.
After the construction of the Dirac bracket structure \cite{Dir}  
we obtain  the   commutators for the fundamental operators together with
 the Heisenberg equations of motion
  (the Hamiltonian operator has no ordering problems)

\begin{mathletters}
\label{16}
\begin{eqnarray}
&&\dot a_n^{\alpha}=-\omega _n M ^{\alpha \beta }b_n^{\beta}\,\,\, ,
 \\
&&\dot b_n^{\alpha}=\omega _n M ^{\alpha \beta }a_n^{\beta}\,\, ,
 \end{eqnarray}
\end{mathletters}
 and their solutions in terms of creation and anihilation operators
 
 \begin{mathletters}
\label{17}
\begin{eqnarray}
\hat{a}_n^{\alpha}(x^0)=\sqrt {\frac{\pi}{2\omega _n}}\,\,\hat
{\Lambda}_{n}^{\alpha}
e^{-i\omega _n x^0}
+\sqrt {\frac {\pi}{2\omega _n}} \,\,\hat{\Lambda}_{n}^{\alpha \dagger}
e^{i\omega _nx^0}, \\
\hat{b}_n^{\alpha}(x^0)=-i\sqrt {\frac{\pi}{2\omega _n}}M ^{\alpha 
\beta}
\,\,\hat
{\Lambda}_{n}^{\beta}
e^{-i\omega _n x^0}\nonumber \\
+i\sqrt {\frac {\pi}{2\omega _n}} M ^{\alpha \beta}\,\,
\hat{\Lambda}_{n}^{\beta \dagger}
e^{i\omega _nx^0}\,\, .
\end{eqnarray}
\end{mathletters}

As in FJ model the problem of the uniqueness of vacuum
(due to the presence of the arbitrary functions $\rho ^{\alpha} $)
 can be solved
by simply defining the field operators as

\begin{eqnarray}
\label{18}
\hat{\Phi}^{\alpha}(x^{0}\,,\,x^{1})\,\,& \equiv & \,\, \hat{\phi}^{\alpha}
(x^{0}\,,\,x^{1})\,
+\, \frac{\hat {\rho}^{\alpha}}{2R}\,\,\,  .
\end{eqnarray}
Using  (\ref{8}) and (\ref{16}) we finally obtain as quantum solutions
\begin{eqnarray}
\label{19}
\hat{\Phi}(x^{0}\,,\,x^{1}) = \nonumber \\ \frac{1}{\sqrt{2\pi}}
\left( \frac{\pi}{R} \right) \sum_{n > 0} \frac{1}{\sqrt{\omega _n}}
\left[ (\delta ^{\alpha \beta}-M ^{\alpha \beta})
\hat{\Lambda}^{\beta}_{n}
e^{-i\omega _n (x^0 - x^{1})}\,\nonumber  \right.\\ +\,\left.
 (\delta ^{\alpha \beta}+M ^{\alpha \beta})
 \hat{\Lambda}^{\beta \dagger}_{n}
  e^{i\omega
_n(x^0+x^{1}) }  \right.\nonumber \\
 + (\left.\delta ^{\alpha \beta}+M ^{\alpha \beta})
\hat{\Lambda}^{\beta}_{n}
e^{-i\omega _n (x^0 + x^{1})}\, \nonumber \right.\\+ \left. \,
 (-\delta ^{\alpha \beta}+M ^{\alpha \beta})
 \hat{\Lambda}^{\beta \dagger}_{n}
  e^{i\omega
_n(x^0-x^{1}) } \right]\,\,\, ,
\end{eqnarray}
with equal-time commutation relations
(taking the limit $ R \rightarrow \infty$)

\begin{equation}
\label{20}
\mbox{} [\hat{\Phi}^{\alpha}(x^{0},x^{1})\,,\,\hat{\Phi}^{\beta}
(x^{0},y^{1})] \,
=\,-\frac{i}{2} M ^{\alpha \beta}\epsilon
(x^{1}\,-\,y^{1})\,\,\, .
\end{equation}
These field operator solutions obey exactly the equations of motion found by
 Tseytlin \cite{Tsey}

 \begin{equation}
 \partial _0\hat \Phi _1 =\partial _1 \hat \Phi _2 \,\,\,\,\,\,
\partial _0\hat \Phi _2 =
\partial _1 \hat \Phi _1 \,\,\, ,\label{21}
\end{equation}
showing that they are the correct gauge-independent quantum field solutions
for this model.

Now we switch our attention to the Poincar\'e algebra of the energy-momentum
tensor. The gauge invariant and symmetric quantum
energy momentum tensor components are
 
\begin{mathletters}
\label{22}
\begin{eqnarray}
&&\hat{\Theta}^{0 0}\,= \hat{\Theta}^{11}\, = \frac{1}{2}\,: (\partial_{1}
 \hat{\Phi ^{\alpha}})
(\partial_{1} \hat{\Phi} ^{\alpha}) : \\
&&\hat{\Theta}^{0 1}\,= \hat{\Theta}^{1 0}=\,:
 (\partial_{1} \hat{\Phi ^{1}})
(\partial_{1} \hat{\Phi} ^{2}) : \,\, .
\end{eqnarray}
\end{mathletters}
Using the quantum solutions (\ref{19}) we obtain, without anomalies,
 the following Virasoro algebra

\begin{mathletters}
\label{23}
\begin{eqnarray}
\mbox [\hat{\Theta}^{00}(x)\,,\, \hat{\Theta}^{00}(y)]\,&
 =
& \,i\,\left(\hat{\Theta}^{01}(x)\,+\,\hat{\Theta}^{01}(y)
\right) \partial_{x^{1}} \delta  \, ,  \\
\mbox [\hat{\Theta}^{00}(x)\,, \, \hat{\Theta}^{01}(y)]\,&
 =
& \,i\,\left(\hat{\Theta}^{00}(x)\,+\,\hat{\Theta}^{00}(y)
\right) \partial_{x^{1}} \delta  \, , \\ 
\mbox [\hat{\Theta}^{01}(x)\,,\, \hat{\Theta}^{01}(y)]\,&
 =
& \,i\,\left(\hat{\Theta}^{01}(x)\,+\,\hat{\Theta}^{01}(y)
\right) \partial_{x^{1}} \delta  \, ,
\end{eqnarray}
\end{mathletters}
that leads to the usual Poincar\'e algebra for the integrated currents  
showing the complete consistency of this formulation under periodic
 boundary conditions.

In the anti-periodic case the Tseytlin model behaves as a purely second-class
constraints system, so no gauge-dependent  modes appear. The Dirac program
can be straightfowardly followed and the excitations obey
 strictly
(\ref{21}) with the energy momentum components consistently obeying
 (\ref {23}).

A global supersymmetric extension can be constructed for this model. Following
the same prescription used in \cite{Tsey} for the bosonic part we find as
extended lagrangian density

\begin{eqnarray}
\label{24}
{\cal L} = \frac{1}{2} \left[ \partial_{0} \phi ^1 \partial_{1} \phi ^2+
\partial_{0} \phi ^2 \partial_{1} \phi ^1 - \partial_{1} \phi ^1 
\partial_{1} \phi ^1 - \partial_{1} \phi ^2 
\partial_{1} \phi ^2 \right. \nonumber   \\
 \left. + i\psi ^{1}\partial _0\psi ^{1}
 + i\psi ^{2}\partial _0\psi ^{2} -i\psi ^{1} 
\partial _1 \psi ^{2}
-i\psi ^{2} 
\partial _1 \psi ^{1}
\right]\,\,\, ,
\end{eqnarray}
that is invariant under the following supersymmetric transformations
\begin{equation} 
\label{25}
\delta \psi ^{\alpha}=i\epsilon M ^{\alpha \beta }\partial _1 
\phi ^{\beta}
\,\,\,\,\,\,\,\,\,\, \delta \phi ^{\alpha}=\epsilon \psi ^{\alpha}\,\,\, ,
\end{equation}
with $\epsilon $ being a constant Grassman parameter. This is exactly
 the 2-dimensional counterpart of the 4-dimensional Schwarz-Sen
 supersymmetric model,  and two succesive susy
 transformations give accordingly a Poincare translation.
 
  The dynamics of this extension is fully consistent at
  the
  classical level. The Hamiltonian analysis gives for the momenta
  
  \begin{equation}
  \pi ^1(x)=-i \psi ^1(x)\,\,\,\,\,\,\,\, \pi ^2(x)=-i \psi ^2(x)   
    \, ,    \end{equation}
  which are second class constraints. The canonical Hamiltonian can also
  be easily constructed, together with the equations of motion (after
  the Dirac brackets construction)
  
  \begin {equation}
H_c=i\psi ^1\partial _1\psi ^2+i\psi ^2 \partial _1 \psi ^1
\, ,\end {equation}

\begin{equation}
\partial _0\psi ^{\alpha }(x)=\{\psi ^{\alpha} ,H\}_D=M_{\alpha \beta }\partial _1
 \psi ^{\beta}(x)  
\, ,\end{equation}
which are exactly equal to the Euler-Lagrange equations, as expected.

The fundamental Dirac brackets between fields are found to be

\begin{equation} 
\{\psi ^i(x),\psi ^i(y)\}_D=i\delta (x-y)\,\,\,\, i=1,2 \, ,
\end{equation}
the others being zero,
 establishing the conditions for a canonically quantized version.

 Following the results obtained for the bosonic model   a  quantum analysis
,searching anomaly terms in the susy algebra,  would be
   of interest. This
  is under investigation and is going to be described in a following work.
\vspace{1cm}

{\bf Acknowledgements}

\vspace{0.5cm}
The authors would like to thank  O. Piguet and N. Berkovits for useful
 discussions. F.P.D. would like to thank the Physics Departments
 of UFPR (Curitiba) and UFES (Vit\'oria)  for hospitality and financial 
 support. C.P.C. would like to thank CNPq (Brazil) for financial support.

\end{document}